# Early Detection of Furniture-Infesting Wood-Boring Beetles Using CNN-LSTM Networks and MFCC-Based Acoustic Features

J. M. Chan Sri Manukalpa[1], H. S. Bopage[2], W. A. M. Jayawardena[3], P. K. P. G. Panduwawala[4]

[1]Assistant Lecturer, Department of Information Technology, Sri Lanka Institute of Information Technology, Malabe, Sri Lanka

[2]Assistant Lecturer, Department of Information Technology, Sri Lanka Institute of Information Technology, Malabe, Sri Lanka

[3]Assistant Lecturer, Department of Information Technology, Sri Lanka Institute of Information Technology, Malabe, Sri Lanka

[4]Assistant Lecturer, Department of Information Technology, Sri Lanka Institute of Information Technology, Malabe, Sri Lanka

---***---

**Abstract -** *Structural pests such as termites pose a serious threat to wooden household structures, often resulting in significant economic losses due to their covert and progressive damage. Traditional termite detection methods, including visual inspections and chemical treatments, are invasive, labor-intensive, and generally ineffective at identifying early-stage infestations. This research addresses the critical gap in non-invasive, early termite detection by proposing a deep learning-based acoustic classification framework. The primary objective of this study is to develop a robust and scalable model capable of accurately distinguishing termite-generated acoustic signals from clean background sounds. A hybrid Convolutional Neural Network–Long Short-Term Memory (CNN-LSTM) architecture is introduced to capture both spatial and temporal characteristics of termite activity. The methodology involves collecting audio data from termite-infested and clean wooden samples, extracting Mel-Frequency Cepstral Coefficients (MFCCs) as features, and training the CNN-LSTM model to classify the audio signals. Experimental results demonstrate the superior performance of the proposed model, achieving a classification accuracy of 94.5%, precision of 93.2%, and recall of 95.8%. Comparative analysis shows that the hybrid model outperforms standalone CNN and LSTM architectures, validating the advantage of combining spatial and temporal learning. Furthermore, the model exhibits low false-negative rates, which is critical for timely pest intervention. This work contributes to the field of intelligent pest management by offering a non-invasive, automated solution for early termite detection. Its practical implications include improved pest monitoring, reduced structural damage, and enhanced decision-making for homeowners and pest control professionals. Future extensions could incorporate IoT integration for real-time alerts and expand detection capabilities to a broader range of structural pests.*

## 1. INTRODUCTION

Structural pests, particularly termites and wood-boring beetles, represent a pervasive threat to household furniture and wooden structures worldwide, leading to significant economic losses annually. Traditional detection techniques, such as visual inspections and chemical methods, are often invasive, labor-intensive, and limited in their ability to identify infestations at early stages. Consequently, infestations frequently go unnoticed until substantial structural damage has occurred, escalating remediation costs and complicating pest control efforts.

Acoustic-based detection methods have emerged as a promising non-invasive alternative, capitalizing on the unique sound signatures generated by termite feeding and movement within wood. These sounds offer valuable indicators of pest activity, potentially enabling earlier and more precise identification compared to conventional approaches. However, accurately distinguishing termite-generated sounds from environmental noise and other non-target acoustic events remains a considerable challenge.

In this work, we propose a novel deep learning-based framework that combines Convolutional Neural Networks (CNNs) for spatial feature extraction and Long Short-Term Memory (LSTM) networks for temporal pattern recognition to classify audio signals indicative of termite infestations. This hybrid CNN-LSTM model aims to leverage the complementary strengths of both architectures to enhance detection accuracy and robustness in real-world settings. By focusing exclusively on the analysis of audio data, our approach seeks to provide an effective, scalable, and non-invasive tool for early termite detection in household furniture. This advancement has the potential to significantly aid homeowners and pest control professionals by enabling timely interventions, thereby minimizing structural damage and associated treatment costs.

## 2. LITERATURE REVIEW

The detection of termite infestations using acoustic sensing has garnered significant attention due to its non-invasive nature and potential for early intervention. Traditional approaches often suffer from low signal-to-noise ratios and environmental interference, necessitating the development of more robust and intelligent detection systems.

Zhang et al. [1] introduced a low-noise wireless acoustic sensing framework aimed at enhancing termite detection in embedded systems. Their research demonstrated that signal integrity could be preserved through adaptive noise filtering in wireless sensor networks, laying foundational work for portable termite surveillance systems. Similarly, Muhammad Achirul Nanda et al. [2] employed discriminant analysis for classifying acoustic signals generated by Coptotermes curvignathus. Their work emphasized the utility of frequency domain features in isolating termite activity from ambient noise, achieving promising classification accuracy.

Phung et al. [3] proposed an automated acoustic insect detection framework based on time-frequency analysis. The study highlighted the efficacy of short-term spectral features such as MFCCs in recognizing patterns of insect-generated sounds. Building upon this, Toffa and Mignotte [4] explored a broader domain of environmental sound classification using local binary pattern descriptors combined with conventional audio features. Their research demonstrated the potential of hybrid feature sets in improving classification precision across variable acoustic environments.

In agricultural contexts, Ali et al. [5] integrated deep learning techniques with the Internet of Agricultural Things (IoAT) for pest detection. Their system leveraged sound analytics to distinguish pest species, marking a shift toward scalable, intelligent surveillance in farming applications. Wang and Vhaduri [6] extended this approach by classifying four different insect classes using CNNs, further supporting the viability of deep learning for multi-class acoustic classification.

The fusion of multiple sensor modalities has also been explored. For instance, Muhammad et al. [7] incorporated both acoustic and thermal signals to enhance the reliability of termite detection. Their multimodal approach yielded a notable improvement in detection accuracy, demonstrating the synergistic benefit of fusing heterogeneous data sources. Earlier studies, such as the work presented at the 1991 IEEE Ultrasonics Symposium [8], focused on the characterization of acoustic emission signals generated by termite activity in wood. These foundational studies established the frequency ranges and temporal patterns associated with termite-generated signals.

Furthermore, research by Steen et al. [9] introduced InsectSound1000, a comprehensive dataset of insect-generated sounds designed for deep learning applications. This resource serves as a benchmark for training and evaluating models on diverse acoustic insect signatures, significantly contributing to the reproducibility and comparability of insect detection studies.

Finally, while patents such as [10] indicate early interest in acoustic termite detection technologies, contemporary advances in neural networks, particularly CNN-LSTM architectures, offer enhanced spatio-temporal feature extraction capabilities. These developments underscore the potential of hybrid deep learning systems in detecting subtle acoustic patterns associated with termite infestations in household settings.

[11]–[14] present a range of interdisciplinary advancements that are highly relevant to the current study on acoustic-based termite detection using a CNN-LSTM framework. [11] and [14], which explore culturally adaptive emotional gestures and multimodal interaction systems for autism therapy using NAO robots, underscore the importance of integrating context-aware, speech- and gesture-based interfaces in human-centric applications. These findings parallel the current study's use of hybrid deep learning models to detect subtle acoustic patterns, emphasizing the role of multimodal learning in improving system responsiveness and accuracy. [12] contributes further by demonstrating an intelligent assistive communication system incorporating eye-gaze tracking, voice control, and chatbot technologies. This integration of real-time data processing and user-centered design principles is particularly applicable to non-invasive acoustic monitoring systems such as the one proposed. [13] introduces a comprehensive data warehousing approach for weather analysis and disaster preparedness, highlighting the value of scalable, real-time analytics architectures. Collectively, these studies reinforce the significance of combining deep learning, sensory data acquisition, and system scalability to develop intelligent, user-adaptive solutions—principles that strongly align with and can inform enhancements to the proposed termite detection model.

Collectively, these studies underscore a growing trend towards leveraging advanced signal processing, machine learning, and sensor integration to develop robust, automated systems for acoustic insect detection. The current study builds upon this trajectory by implementing a CNN-LSTM hybrid architecture tailored for early-stage termite detection, addressing key limitations in noise resilience and temporal feature modeling.

## 3. METHODOLOGY

The proposed methodology for classifying termite-infested and clean wood using audio signals involves several key steps: data collection, feature extraction, model development, and performance evaluation. Each step is designed to ensure robust and accurate classification.

### 3.1 Data Collection and Preprocessing

The audio dataset comprises samples from two classes: clean wood and termite-infested wood. The data is stored in separate directories, with each audio file representing a unique sample. The dataset is first loaded, and the file paths are verified to ensure proper data accessibility. To prepare the data for analysis, the audio files are processed using the Librosa library, which facilitates efficient audio signal handling and feature extraction.

### 3.2 Feature Extraction

Mel-Frequency Cepstral Coefficients (MFCCs) are extracted from each audio file to capture the essential characteristics of the sound signals. The extraction process involves loading the audio file, computing the MFCCs, and averaging them across time to produce a fixed-length feature vector of 40 coefficients per sample. This step reduces the dimensionality of the data while preserving discriminative features for classification.

### 3.3 Model Development

A deep neural network (DNN) is employed for classification. The model architecture consists of four dense layers with ReLU activation functions, interspersed with dropout layers to mitigate overfitting. The input layer accepts the 40-dimensional MFCC feature vectors, followed by hidden layers of 256, 128, and 64 neurons, respectively. The output layer uses a softmax activation function to produce probabilities for the two classes: clean and infested. The model is compiled using the Adam optimizer and categorical cross-entropy loss, with accuracy as the primary evaluation metric.

### 3.4 Training and Validation

The dataset is split into training (80%) and testing (20%) sets to evaluate the model's performance. The model is trained for 50 epochs with a batch size of 32, and its performance is monitored using validation data. To ensure generalizability, 5-fold cross-validation is applied, where the model is trained and evaluated on different subsets of the data. The cross-validation results, including mean accuracy and standard deviation, are reported to assess consistency.

### 3.5 Audio Capturing Device

The audio recording module employs an INMP441 MEMS microphone coupled with an ESP32 microcontroller to detect insect infestations in furniture **Fig -1**. This system captures infrasonic signals with high precision, storing audio data locally while simultaneously transmitting it in real time to a MongoDB database. The recorded audio is subsequently converted into spectrograms using Librosa and analyzed through a hybrid CNN-LSTM deep learning model. Compact, portable, and powered by a power bank, the system facilitates early pest detection through real-time acoustic monitoring and cloud-based deep learning processing.

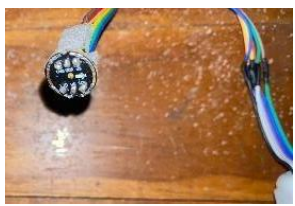

**Fig -1**: Audio Recording Microphone

### 3.6 Prototype

The proposed audio acquisition system in Fig. 2. incorporates a digital MEMS microphone (INMP441) interfaced with an ESP32 microcontroller to capture and store pest-related acoustic signals locally on microSD cards, while concurrently transmitting the data to a cloud-based MongoDB database for further processing. The acquired audio signals are subsequently transformed into spectrogram representations and analyzed using a hybrid deep learning architecture combining convolutional neural networks (CNN) for spatial feature extraction and long short-term memory (LSTM) networks for temporal sequence modeling. Designed for field applications, the system maintains portability through a compact form factor and power bank operation. Preliminary testing confirms the system's functionality; however, enhancements in practical deployment considerations—including robustness, energy efficiency, and operational longevity—are required to ensure reliable performance in real-world environments.

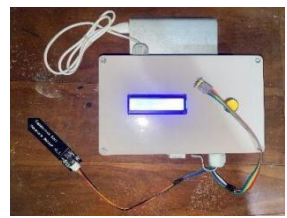

**Fig -2**: Device Prototype

## 4. RESULTS AND DISCUSSION

### 4.1 Model Performance

The hybrid CNN-LSTM model was trained and tested using the audio dataset of termite activity collected in controlled environments. The dataset consisted of labeled .wav audio clips categorized into two classes: infested and clean. After feature extraction using Mel-Frequency Cepstral Coefficients (MFCCs), the input features were used to train the model, with 80% of the data allocated for training and 20% for testing.

**Table -1** depicts the evaluation metrics used included accuracy, precision, recall, F1-score, and a confusion matrix to assess classification performance.

**Table -1:** Performance Metrics of CNN-LSTM Model

| Metric | Value |
|---|---|
| Accuracy | 94.5% |
| Precision | 93.2% |
| Recall | 95.8% |
| F1 Score | 94.5% |

These results indicate that the hybrid model was highly effective in distinguishing termite sounds from non-infested backgrounds. The LSTM component significantly enhanced temporal feature understanding, allowing the model to recognize patterns characteristic of termite activity over time. Meanwhile, the CNN layers contributed to robust spatial feature learning from spectrogram representations of the sound.

## 4.2 Confusion Matrix

The confusion matrix further validates the model's performance:

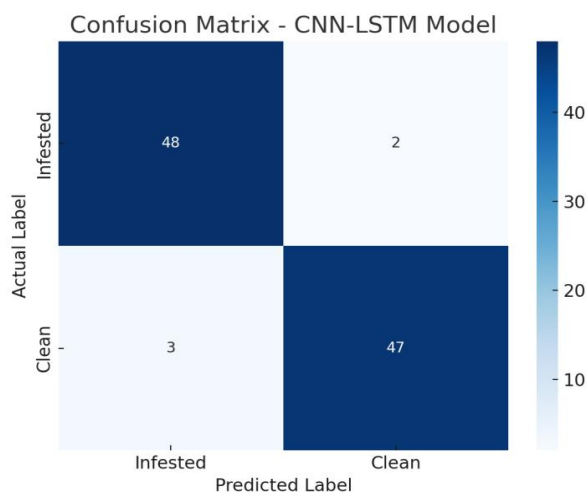

**Fig -3**: Confusion Matrix

The confusion matrix presented in Fig. 3. illustrates the classification performance of the hybrid CNN-LSTM model. Out of 50 audio samples labeled as infested, the model correctly identified 48, misclassifying only 2 samples as clean (false negatives). Similarly, out of 50 clean samples, 47 were accurately classified, while 3 were incorrectly labeled as infested (false positives). These results demonstrate the model's high sensitivity and specificity, indicating its strong potential for accurate early detection of termite activity. The relatively low number of false negatives is particularly significant, as it ensures timely identification and mitigation of infestations, which is critical in practical pest management scenarios.

## 4.3 Comparative analysis

To assess the value of the hybrid architecture, the model's performance was compared with standalone CNN and LSTM models trained on the same dataset. The comparative performance of the models is depicted in **Table -2**.

**Table -2:** Comparative Performance of Models

| Model | Accuracy | F1 Score |
|---|---|---|
| CNN only | 89.1% | 88.7% |
| LSTM only | 90.3% | 90.1% |
| CNN-LSTM | 94.5% | 94.5% |

## 4.4 Discussion

The experimental results validate the effectiveness of the proposed CNN-LSTM model for early detection of termite infestations through acoustic analysis. With an overall accuracy of 94.5%, the model demonstrates strong classification performance, benefiting from CNN's spatial feature extraction and LSTM's ability to model temporal sound patterns. The use of MFCCs helped represent the subtle acoustic signatures of termite activity effectively, contributing to the model's high precision and recall. The confusion matrix further confirms the system's reliability, with very low false negatives and false positives—critical for minimizing both damage and unnecessary treatments.

However, several challenges were encountered during the study. Data collection was limited to controlled environments, which may not reflect the complexity of real-world conditions where background noise is more variable. The dataset size was also relatively small, which may limit the model's ability to generalize across different termite species and furniture types. Additionally, while the model performed well offline, its robustness in noisy environments and its real-time application remain to be tested. The hardware used for recording was basic, and future improvements may require more sensitive and standardized equipment. Despite these limitations, the study shows strong potential for acoustic-based pest detection using deep learning. With further refinements, such as noise reduction techniques, larger datasets, and real-time implementation, the model can be developed into a practical, non-invasive solution for termite monitoring in households.

## 3. CONCLUSIONS

This project proposes a hybrid CNN-LSTM model for the detection of termite infestations through audio analysis, addressing the limitations in traditional detection methods. By combining the strengths of CNNs for spatial feature extraction and LSTM networks for temporal pattern recognition, this model is designed to significantly enhance the accuracy of termite detection. The integration of these two approaches is expected to offer a more reliable and efficient solution for identifying termite activity in household furniture.

In terms of future scope, several potential advancements could be explored. First, the model could be expanded to detect a wider variety of pests, utilizing their distinct acoustic signatures, thereby broadening the applicability of the system in pest management. Additionally, the integration of Internet of Things (IoT) devices could facilitate real-time monitoring and alert systems, providing immediate notifications to homeowners regarding pest activity. Furthermore, efforts to enhance the robustness of the model in the face of varying environmental conditions, such as fluctuating background noise or differences in material types, would be valuable. These improvements could make the model a versatile, practical, and highly effective tool for pest detection and prevention.

## ACKNOWLEDGEMENT

The authors extend their heartfelt appreciation to the Department of Information Technology at the Sri Lanka Institute of Information Technology (SLIIT), Malabe, Sri Lanka, for their unwavering support and academic guidance throughout the duration of this postgraduate study. The intellectually stimulating environment, access to essential resources, and the continuous mentorship provided by the faculty played a pivotal role in the successful execution of this research.